\documentclass[reprint,nofootinbib,preprintnumbers,amsmath,amssymb,onecolumn,unsortedaddress,superscriptaddress,twocolumn]{revtex4-1}
\usepackage[utf8]{inputenc}
\usepackage[normalem]{ulem}
\usepackage[table]{xcolor}
\usepackage[rightcaption]{sidecap}
\usepackage{graphicx}
\usepackage{booktabs,caption}

\begin{document}
\title{Coherent Collections of Rules Describing Exceptional Materials \\ Identified with a Multi-Objective Optimization of Subgroups}
\author{Lucas Foppa*}
\affiliation{The NOMAD Laboratory at the Fritz Haber Institute of the Max Planck Society, Faradayweg 4-6, D-14195 Berlin, Germany}
\author{Matthias Scheffler}
\affiliation{The NOMAD Laboratory at the Fritz Haber Institute of the Max Planck Society, Faradayweg 4-6, D-14195 Berlin, Germany}

\begin{abstract}
Useful materials are often statistically exceptional and they might be overlooked by AI models that attempt to describe all materials simultaneously. These global models perform well for the majority of (useless) materials, but they do not necessarily capture the useful ones. 
Subgroup discovery (SGD) identifies \textit{rules} describing subsets of materials (SGs) associated to exceptional values, e.g., high values, of a materials property of interest. Thus, SGD can better capture exceptional materials compared to most widely used AI techniques. Previous works focused on the SG that maximizes an objective function that establishes one tradeoff between the size of the SG and the exceptionality of the distribution of property values in the SG. However, this optimization does not give a unique solution, but many SGs typically have similar objective-function values. 
Here, we identify a "Pareto region" of SGs presenting a multitude of size-exceptionality tradeoffs. The approach is demonstrated by the learning of rules describing perovskites with high bulk modulus. These rules are used to screen a large space of perovskites and to efficiently identify materials with bulk modulus up to 13\% higher than the highest value of the training set.  
\end{abstract}
\date{\today}
\maketitle

\section{Introduction}
Materials serve as the cornerstone of critical economic sectors and they play a pivotal role in driving the transition to a sustainable economy and to renewable energy.\cite{Lewis-2006, Chu-2012} Thus, there is an urgent need for the discovery of appropriate and more efficient materials. However, the materials that are useful for a given application are typically statistically exceptional. 
These materials might present, for instance, extremely high values of a materials property compared to other known compounds. Exceptional materials are very few compared to the practically infinite space of possible materials, which remains largely unknown.\cite{Davies-2016, Schrier-2023} Artificial intelligence (AI) has been increasingly applied in materials science and engineering.\cite{Ramprasad-2017, Schmidt2019, Raabe-2023, Bauer-2024} Indeed, AI might describe materials properties and functions governed by intricate mechanisms better than previous theoretical and computational approaches because it targets correlations and does not assume a single underlying physical model.\cite{Foppa-2021} Thus, AI holds the potential to accelerate the exploration of the immense materials space, leading to the discovery of new materials. However, capturing exceptional materials is a challenging task for most widely used AI methods. 

AI methods often fail in describing the exceptional materials firstly because the training data is typically not well distributed over (or representative of) the huge, unknown materials space. Therefore, interpolation schemes are unable to generalize to potentially interesting portions of the materials space that were disregarded in the training data.\cite{Meredig-2018, Kauwe-2020, Muckley-2023, Li-2023} This issue can be alleviated by AI approaches that can better extrapolate compared to methods that are inherently interpolative.\cite{Muckley-2023, Wang-2019} Besides, AI model training can be combined with the systematic acquisition of new data corresponding portions of the materials space that were not covered by the initial training data using sequential-learning approaches such as active learning or Bayesian optimization.\cite{Shahriari-2016,Zhang-2023, Siemenn-2023, Biswas-2024} However, the efficiency of sequential learning often relies on the quality of uncertainty estimates, which is in some cases problematic.\cite{Hirschfeld-2020, Palmer-2022, Tran-2020} 
A second key reason that can explain the inability of current AI approaches to capture exceptional materials is the focus on global models. These models attempt to describe all materials simultaneously. They are obtained by optimizing an objective (loss) function that reflects the average performance, e.g., the mean prediction error. Thus, global models are designed to perform well in average for the majority of (uninteresting) materials, but do not necessarily perform well for exceptional ones.\cite{Borg-2023} The objective functions can be adapted to give more importance to the description of specific property values, e.g., high values.\cite{Serraino-2013} However, different groups of materials could operate according to different mechanisms. This might render a global description not only inaccurate, but also inappropriate.  

\begin{figure*}[ht!]
\centering
\includegraphics[width=17.5cm]{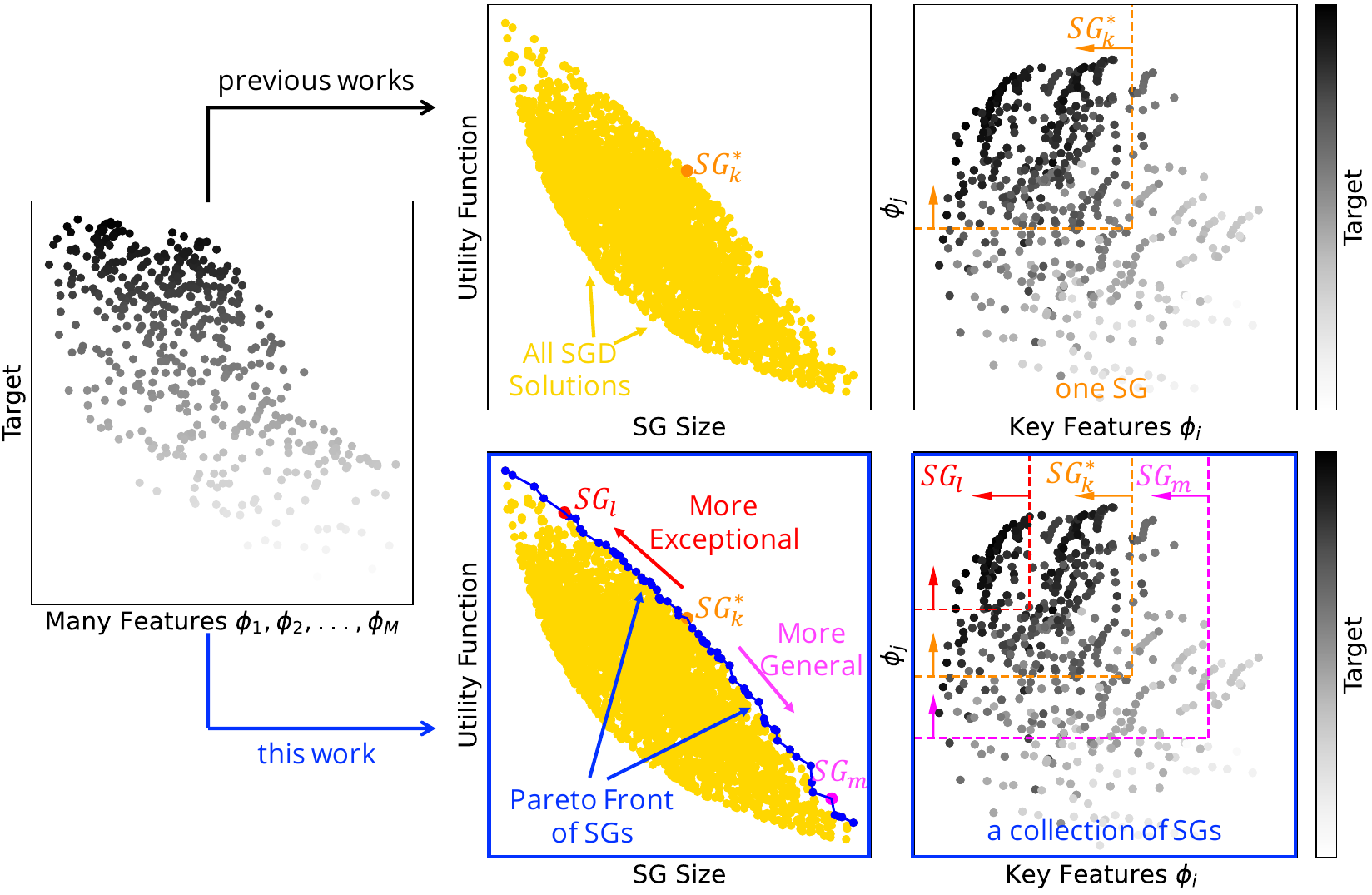}
\caption{Subgroup discovery (SGD) identifies subsets of materials (SGs) that are outstanding with respect to a certain materials property, the target of interest. These subsets are described by rules that typically constraint key features characterizing the materials and mechanisms governing the target, out of many initially offered features. The SGs are obtained by maximizing an objective function that is a product of the (relative) SG size and the utility function. These two terms reflect the generality and the exceptionality of the SG, respectively. Top: Previous works focused on the one SG that maximizes the objective function, denoted as $SG_k^*$. Bottom: In this work, we identify a collection of SGs containing, e.g., $SG_l$, $SG_k^*$, and $SG_m$, presenting a multitude of tradeoffs between the SG size and utility function.}
\label{fig:scheme_SGD}
\end{figure*}

Alternative AI methods for materials discovery include strategies based on the similarity among materials\cite{Seko-2018,Kuban-2022} or among their constituents, e.g., ions in solids,\cite{Jia-2022} and the subgroup-discovery (SGD)\cite{Wrobel-1997,Friedman-1999} approach. 
In particular, SGD tackles the limitations of global descriptions and it has the potential to better capture exceptional materials. Indeed, SGD has been recently put forward in materials science.\cite{Goldsmith-2017,Boley-2017,Sutton-2020,Li-2021,Foppa-TopCatal-2022,Foppa-ACSCatal-2022} 
It is a supervised, descriptive rule-induction\cite{Novak-2010} technique and it identifies subsets of a dataset associated to exceptional values of a target quantity of interest, for instance a materials property or performance indicator. Crucially, SGD identifies these subsets (SGs) of data along with the descriptions of these subsets, referred to as \textit{rules}. The SGD analysis starts with the choice of many features that relate to possibly relevant mechanisms governing the target materials property.
Then, SGD creates a number of statements about the features that are satisfied only for part of the dataset. These statements are, for instance, inequalities constraining the values of the features. Finally, a search algorithm\cite{Boley-2011,Boley-2012,Grosskreutz/2008} identifies the combination of typically few statements that results in a SG that maximizes an objective function. This objective (or quality) function is a product of the relative SG size and the so-called \textit{utility function}. The relative SG size is the fraction of data points that satisfies the statements describing the SG. The higher the relative SG size, the more general the description. The utility function quantifies the ''exceptionality" of the distribution of target values in the SG with respect to the entire dataset. The positive mean shift is one example of an utility function often utilized when the target values of interest are high. This utility function measures the shift of the mean value of the target in the SG with respect to mean value of the target in the entire dataset. The higher the value of the positive-mean-shift utility function, the higher the target values in the identified SG. Thus, such utility function favors the identification of SGs associated with high target values. We will discuss this utility function in more details in the Results section. We will also discuss a second example of utility function, namely the Jensen-Shannon divergence between the distributions of target values in the SG and in the entire dataset.
The SG rules typically constraint the values of only few, key features, out of the many initially offered ones. Thus, SGD learns a (low-dimensional) \text{representation}. 
The SGD approach is illustrated in Fig.~\ref{fig:scheme_SGD} (Top). 
The aim of SGD is to find descriptions of portions of the materials space that are exceptional. 
Thus, it accepts that the mechanisms governing the materials' property might vary across the materials space and that not all of these mechanisms need to be described for the discovery of useful materials. Indeed, SGD was used to identify rules associated with high-performance materials even based on datasets dominated by low-performance situations.\cite{Foppa-ACSCatal-2022} 
Additionally, SGD is an exploratory analysis that can identify unexpected patterns and anomalies. 
We note that SGD is significantly different from clustering techniques because it does not aim at describing the entire dataset, it is a supervised approach, and it explicitly identifies rules indicating why the data points belong to the SG. Clustering is an unsupervised technique that groups data points into clusters based on similarity, without considering any target quantity. Besides, clustering does not explicitly identify why the data points are clustered together.


Previous SGD works\cite{Goldsmith-2017,Boley-2017,Sutton-2020,Li-2021} focused on the identification of the SG (and rules) that maximizes the objective function, as illustrated in Fig.~\ref{fig:scheme_SGD} (top). 
However, the SG that maximizes the objective function does not reflect all possible tradeoffs between relative SG size and utility function that could be relevant for a given application. 
Additionally, the definitions of some utility functions assume that the distributions of target values in the entire dataset and in the SG are appropriately characterized by one single summary-statistics value, such as the mean value in the case of the positive-mean-shift utility function. However, this assumption may be questioned in in materials science, as the distributions can significantly deviate from the normal distribution. Indeed, distributions related to materials can be skewed or even bimodal. Finally, utility functions often assume that the summary-statistics values appropriately reflect the huge, unknown materials space, e.g., the mean value in the dataset is a good approximation for the mean value in the entire materials space. This assumption does not hold if the datasets are created according to certain selection biases.  Thus, the datasets can be highly unbalanced compared to the materials space.  

In this manuscript, we introduce a multi-objective optimization of SGs that identifies \textit{coherent collections} of SGs and rules at the "Pareto region" of optimal SGD solutions. These SGs present a multitude of tradeoffs between the relative SG size and the utility function, the two conflicting objectives in SGD. This concept is schematically shown in Fig.~\ref{fig:scheme_SGD} (bottom). 
The multi-objective optimization of SGs is demonstrated for the identification of $AB$O$_3$ perovskites with high bulk modulus as an example of a target. We compare the SGs obtained with two different utility functions, the positive mean shift and the cumulative Jensen-Shannon divergence. The latter does not make assumptions on the shape of the distributions of target values. 
We also analyze the sensitivity of the results with respect to offered set of features. Finally, we exploit the rules trained on a dataset of 504 single $AB$O$_3$ perovskites to identify high-bulk-modulus perovskites out of a candidate space of 12,096 single $AB$O$_3$ and double $A_2BB'$O$_6$ perovskites.
Our results show that rules focusing on perovskites with high bulk modulus do not necessarily correspond to the single SG which maximizes the objective function, but they are better, systematically derived with the Pareto-region concept. These rules identify perovskites of the candidate space that present bulk modulus up to 13\% higher than the highest value of the training set. 

\section{Approach for Identifying the Pareto Region of SGD Solutions}
The SGD approach is based on a dataset of materials, which we denote $\widetilde{P}$. This dataset is part of the huge materials space, the full population, $P$. Each material of the full population is associated with a set of features, physical parameters that are potentially related with a target quantity of interest $y$, for instance a materials property. The target of interest is only known for the materials in in dataset. SGD starts by generating a number of statements about the features. Each statement is only verified for a part of the materials in the dataset. For the case of continuous (metric) parameters, the statements are typically inequalities constraining the values of the features to some minimum or maximum thresholds to be determined during the analysis. Then, SGD uses a search algorithm, for instance Monte-Carlo-based\cite{Boley-2011,Boley-2012} or branch-and-bound,\cite{Grosskreutz/2008} to identify conjunctions of statements constructed with the "AND" operator ($\wedge$), that result in SGs that maximize an objective (quality) function $Q$ of the form

\begin{equation}\label{eq:quality_function_alpha}
    Q(SG,\widetilde{P})=\left(\frac{s(SG)}{s(\widetilde{P})} \right) ^{\alpha}(u(SG,\widetilde{P}))^{\beta}.
\end{equation}
Here, $s(SG)$ and $s(\widetilde{P})$ are the sizes of the SG and of the dataset $\widetilde{P}$, respectively, i.e., the number of data points that satisfy the statements defining the SG and the number of data points in the entire dataset. The ratio between the size of the SG and the size of the dataset, $s(SG)/s(\widetilde{P})$, is referred to as the relative SG size. $u(SG,\widetilde{P})$ is the utility function describing how exceptional the distribution of the target in the SG is compared to the entire dataset. The utility function is chosen according to the question to be addressed, and there are many possibilities.\cite{Boley-2017} The positive shift of the mean value of the target in the SG compared to the mean value of target in the entire dataset and the Jensen-Shannon divergence between the distribution of target values in the SG and the distribution of target values in the entire dataset\cite{Nguyen-2015} are two examples of utility functions that we will consider in this work. Finally, $\alpha$ and $\beta$ are tunable parameters controlling the tradeoff between the relative SG size, i.e., the generality of the description, and the utility function, i.e., the exceptionality of the description. Usually, $\alpha=\beta=1$ or $\beta=1-\alpha$, with $\alpha \in [0.1,0.9]$. 
The SGD analysis simultaneously identifies the subsets of data (SGs) and the statements that describe these subsets of data. The latter are the so-called rules. These rules typically depend only on key features, out of all initially offered features. In analogy to genes in biology, these key features might be called \textit{materials genes},\cite{Foppa-2021} as they correlate with the mechanisms governing the materials property. The rules can be exploited to efficiently identify the few exceptional materials in the huge materials space $P$, for which the target property is not known.  

In order to identify the pursued coherent collections of SGs with multiple generality-exceptionality tradeoffs, we first run the SGD algorithm using the objective function of Eq.~\ref{eq:quality_function_alpha}
with $\alpha=\beta=1$. Thus, relative SG size and utility function are given the same importance. 
Then, we collect a number of SGD solutions identified by this algorithm that display high objective-function values. Among these top-ranked SGD solutions, we identify a Pareto front with respect to the two objectives relative SG size and utility function. In multi-objective optimization, a Pareto front is the set of solutions for which no single objective can be improved without deteriorating at least one other objective. Thus, the solutions in the Pareto front reflect an optimal tradeoff between competing objectives. To ensure that no interesting SGD solution is left out, we included in our analysis, not only solutions that are part of the Pareto front, but also solutions within a fixed distance (in this work equal to 0.01) to the Pareto front in the relative SG size-utility function space, i.e., solutions which are \textit{near the Pareto front}. We refer to the solutions at the Pareto front plus the solutions near the Pareto front as \textit{Pareto region}.

\section{Application: Identification of Perovskites with High Bulk Modulus}

\begin{table*}
\centering
\caption{Features used in the SGD analysis to characterize the $AB$O$_3$ perovskites. $^a$Properties of the solid material. $^b$ Properties of free atoms of elements constituting the material. $^c$ Properties of the composition of the material. $^d$ Evaluated using DFT-PBEsol. $^e$ Energy needed per atom to atomize the crystal. $^f$ Defined based on the periodic-table group of the $A$ element and on the charge neutrality of the $AB$O$_3$ composition, i.e., $n_A+n_B=6$.}
\label{tab:Features}
\begin{tabular}{l l l l}
\toprule
Type & Name & Symbol  & Unit\\
\hline
S$^a$ & Equilibrium lattice constant$^d$ & $a_0$ & \AA \\
\hline
S$^a$ & Cohesive energy$^{d,e}$  & $E_0$ & eV/atom \\
\hline
A$^b$ & Radii of the valence-$s$ orbitals of the $A$ and $B$ neutral atoms$^d$ & $r_{s,A}$,$r_{s,B}$   & \AA \\
\hline
A$^b$ & Radii of the valence-$s$ orbitals of the $A$ and $B$ +1 cations$^d$ & $r_{s,A}^{\mathrm{cat}}$,$r_{s,B}^{\mathrm{cat}}$   & \AA \\
\hline
A$^b$ & Radii of the highest-occupied orbitals of $A$ and $B$ neutral atoms$^d$         & $r_{\mathrm{val},A}$ , $r_{\mathrm{val},B}$              & \AA \\
\hline
A$^b$ & Radii of the highest-occupied orbitals of the $A$ and $B$ 1+ cations$^d$        & $r_{\mathrm{val},A}^\mathrm{cat}$, $r_{\mathrm{val},B}^\mathrm{cat}$   & \AA \\
\hline
A$^b$ & Electron affinity of the $A$ and $B$ atoms$^d$  & $EA_A$, $EA_B$                    & eV \\
\hline
A$^b$ & Ionization potential of the $A$ and $B$ atoms$^d$ & $IP_A$, $IP_B$                    & eV \\
\hline
A$^b$ & Electronegativity of the $A$ and $B$ atoms$^d$ & $EN_A$, $EN_B$ & eV \\
\hline
A$^b$ & Kohn-Sham single-particle eigenvalue of the highest-occupied orbital of the $A$ and $B$ atoms$^d$ & $\epsilon_{\mathrm{H},A}$, $\epsilon_{\mathrm{H},B}$ & eV \\
\hline
A$^b$ & Kohn-Sham single-particle eigenvalue of the lowest-unoccupied orbital of the $A$ and $B$ atoms$^d$ & $\epsilon_{\mathrm{L},A}$, $\epsilon_{\mathrm{L},B}$ & eV \\
\hline
A$^b$ & Atomic number of $A$ and $B$ elements & $Z_A$, $Z_B$ & $\mathbb{Z}$ \\
\hline
C$^c$ & Expected oxidation state of the element $A$ and $B$ in the perovskite formula$^f$ & $n_A$, $n_B$  & $\mathbb{Z}$ \\
\botrule
\label{tab:feat}
\end{tabular}
\end{table*}

The identification of coherent collections of SGs and the usefulness of our approach will be demonstrated for the learning of rules describing $AB$O$_3$ perovskites that exhibit high bulk modulus ($B_0$) as an example of a target materials property. Perovskites are a promising materials class\cite{Peña-2001,Travis-2016} for energy-related applications such as photovoltaics and catalysis\cite{Jena-2019, Hwang-2017, Kim-2020} and they have been the subject of a number of AI and machine-learning studies.\cite{Pilania-2016, Bartel-2019, Tao-2021, Ihalage-2021, Gu-2022} The dataset\cite{Foppa-PRL-2022} used to train the SG rules contains 504 cubic perovksites composed by common $A$ elements from the alkali, alkaline-earth, and scandium groups as well as lanthanides. The choice of $B$ elements includes transition-metals as well as main-group elements such as bismuth, antimony, and germanium. We used 24 features characterizing the perovskites (Table~\ref{tab:feat}). Two of the features are properties of the solid perovskite materials, the equilibrium lattice constant ($a_0$) and the cohesive energy ($E_0$). Ten of the features are atomic properties of free atoms of the elements $A$ or $B$, such as orbital radii, ionization potential, and electronegativity. Finally, we included two features that depend on the composition of the material, the expected oxidation states of $A$ and $B$ elements in the compound ($n_A$ and $n_B$, respectively). The bulk modulus and the features (except the atomic numbers of $A$ and $B$, $n_A$, and $n_B$) were calculated using density-functional theory (DFT) with the PBEsol exchange-correlation functional. Calculation details are provided elsewhere.\cite{Foppa-PRL-2022} We note that some of the features in Table~\ref{tab:feat} are correlated with each other. This is not a limitation for SGD. However, the presence of correlated features might result in similar SGs defined by slightly different rules. 

\section{Results}
\subsection*{Collection of SGs Rules Obtained with the Positive-Mean-Shift Utility Function}

\begin{figure*}[ht!]
\centering
\includegraphics[width=17cm]{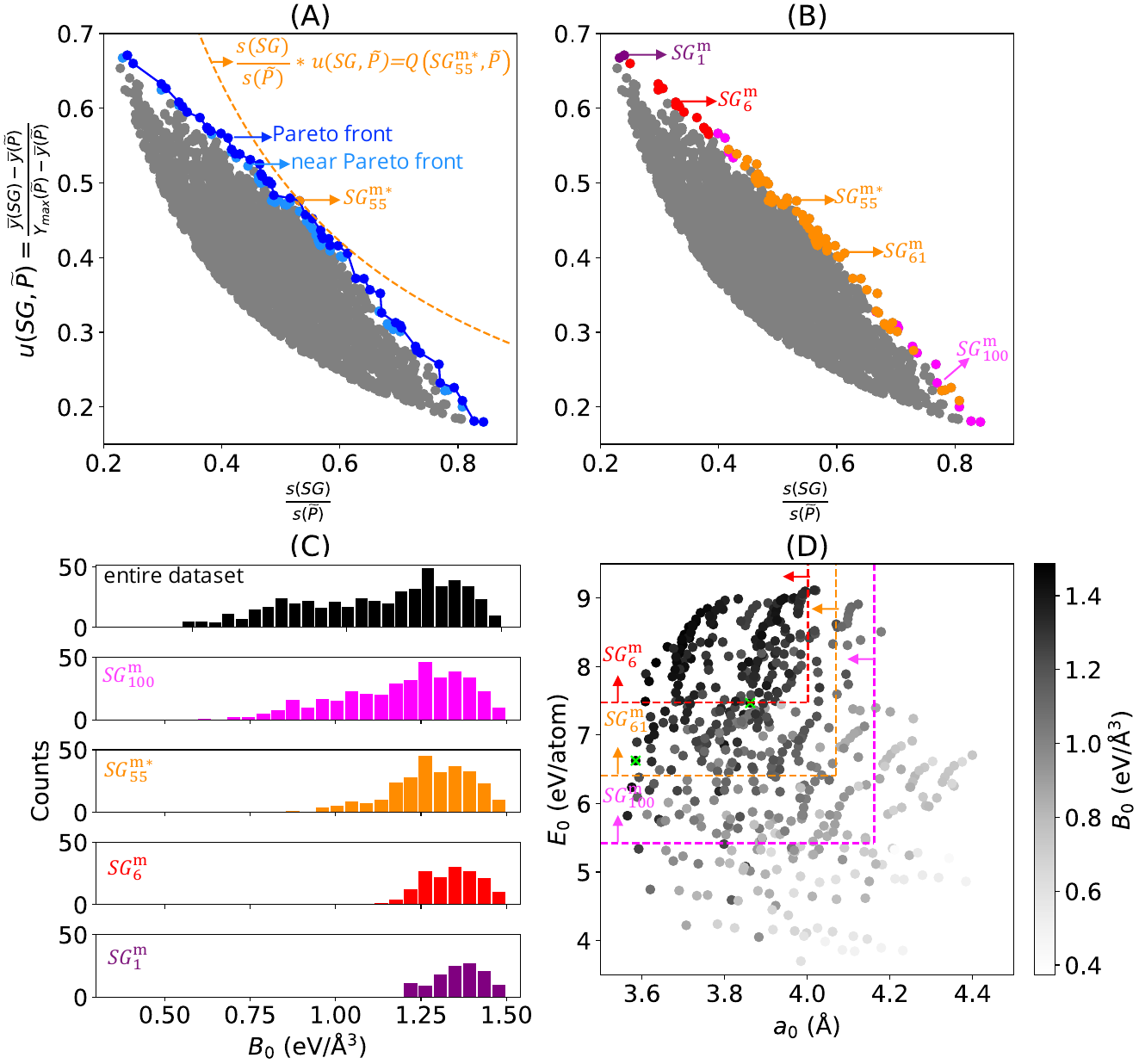}
\caption{Collections of SGs describing perovskites with high bulk modulus ($B_0$) obtained using the multi-objective optimization approach and the positive-mean-shift utility function. Results for the full feature set containing the 24 features of Table I (A) The 5,000 SGD solutions with high values of objective function are shown in grey and Pareto region of SGD solutions is shown in blue. The SG associated to the maximum value of objective function is shown in orange. (B) The 109 SGs of the Pareto region are clustered according to their similarity via  hierarchical clustering. Each cluster is shown in a different color. (C) Distributions of  $B_0$ in the entire training dataset and in some examples of SGs of the Pareto region. The rules associated to these SGs are shown in Table II. (D) SG rules for some examples of SGs of the Pareto region that constrain the values of equilibrium lattice constant ($a_0$) and cohesive energy ($E_0$).}
\label{fig:pos_mean}
\end{figure*}

We start by analyzing the results obtained with the positive-mean-shift utility function, defined as 
\begin{equation}\label{eq:pos_mean}
    u(SG,\widetilde{P})=\frac{\bar{y}(SG)-\bar{y}(\widetilde{P})}{y_{\mathrm{max}}(\widetilde{P})-\bar{y}(\widetilde{P})}.
\end{equation}
Here, $\bar{y}(SG)$ and $\bar{y}(\widetilde{P})$ are the mean values of the distribution of the target in the SG and in the entire dataset, and $y_{\mathrm{max}}(\widetilde{P})$ is the maximum value that the target assumes in the dataset. In our application, the target is the bulk modulus and $y_{\mathrm{max}}(\widetilde{P})$=1.49 eV/$\mathrm{\AA}^3$ for the ScMnO$_3$ perovskite. The utility function of Eq.~\ref{eq:pos_mean} requests that the values of the target within the SG are high with respect to the mean value of the target in the dataset. It assumes that the distributions of target values in the SG and in the entire dataset are properly described by the mean values. 

The 5,000 SGs with the highest objective-function values identified in the analysis using the positive-mean-shift utility function are shown as grey points in Fig.~\ref{fig:pos_mean}(A). The Pareto region is shown in blue in this plot: the 60 and 49 SGs belonging to the Pareto front and to the near-Pareto-front region are displayed in dark and light blue, respectively. This plot shows, in orange, the SG that maximizes the objective function, denoted $SG^{\mathrm{m*}}_{55}$. The curve corresponding to the constant value of $Q(SG^{\mathrm{m*}}_{55},\widetilde{P})$ is shown as a dashed orange line. Note that the we assign the $i$ indices to the SGs of the Pareto region $SG^{\mathrm{m}}_i$ according to increasing values of relative SG size. The star in $SG^{\mathrm{m*}}_{55}$ indicates that this is the SG associated with the maximum objective-function value. The Pareto region contains many SGs with objective-function values close to the maximum at relative SG sizes in the range [0.4,0.6]. Conversely, SGs at the Pareto region with relative sizes lower than 0.4 and higher than 0.6 present relatively lower objective-function values compared to the maximum value.

The Pareto-region concept leads to the identification of not one, but a collection of 109 SGs presenting multiple generality-exceptionality tradeoffs. However, it is unclear how to choose which of these SGs should be considered for a detailed analyzed of physical insights or for materials discovery in larger candidate materials spaces. Many of the SGs of the Pareto region might be similar to each other and they might contain redundant information. 
In order to assess the variability of the SG rules of the Pareto region and to facilitate further analysis of these rules, we established a measure of similarity between SGs and used it to identify \textit{clusters} of SGs containing similar SGs.\cite{Niemann-2017} This analysis is described in details in the Electronic Supporting Information (ESI). In summary, the similarity is assessed by Jaccard indices. These indices consider that the similarity between two SGs is proportional to the overlap of their elements, i.e., to the number of data points that satisfy the rules defining both SGs. Thus, this similarity will be high between SG rules that result in similar selection of materials, even though the rules themselves might be different, e.g., due to correlated key features or due to different thresholds. To obtain the clusters, we applied agglomerative hierarchical clustering.\cite{Nielsen2016} 

The results of the similarity analysis and clustering of the Pareto region of Fig.~\ref{fig:pos_mean}(A) for a chosen number of four clusters is displayed in Fig.~\ref{fig:pos_mean}(B). In this figure, the four identified clusters are displayed in four different colors. In general, the clusters of SGs correspond to different ranges of relative sizes. The cluster shown in orange, for instance, can be related to the SGD solutions with objective-function values close to the maximum at the relative size range of [0.4,0.6]. Interestingly, a small cluster containing only three SGs is identified at low relative sizes. This indicates that these three SG rules are unique compared to the remaining ones. 

We analyzed in more detail one SG per identified cluster. $SG_1^\mathrm{m}$, $SG_6^\mathrm{m}$, $SG_{55}^\mathrm{m*}$, and $SG_{101}^\mathrm{m}$ are examples of SGs belonging to the purple, red, orange, and magenta clusters, respectively. The distributions of bulk-modulus values in the entire dataset, and in the mentioned SGs are shown in Fig.~\ref{fig:pos_mean}(C). As the relative SG size decreases, the SGs of the Pareto region have higher mean bulk-modulus values and narrower bulk-modulus distributions. For the goal of identifying perovskites with extremely high bulk modulus, the rules associated to SGs with low relative SG size and high mean bulk-modulus values, e.g., associated to $SG_1^\mathrm{m}$ or $SG_6^\mathrm{m}$, are useful, since they provide a more focused description. Such SGs would not be detected based on the maximization of the objective function alone. 


Next, we analyzed the rules defining $SG_1^\mathrm{m}$, $SG_6^\mathrm{m}$, $SG_{55}^\mathrm{m*}$, and $SG_{101}^\mathrm{m}$, shown in Table~\ref{tab:Rules}. The rules constrain the values of 3 key features, out of the 24 offered features: equilibrium lattice constant ($a_0$), cohesive energy ($E_0$), and radius of valence-$s$ orbitals of +1 cations (cat) of $B$ element ($r_{s,B}^{\mathrm{cat}}$). In particular, the $a_0$ and $E_0$ values are always constrained to maximum and minimum thresholds, respectively. Thus, perovskites with short lattice constant and high cohesive energy tend to present high bulk modulus. This reflects the inverse relationship of bulk modulus with lattice constant and the direct relationship of bulk modulus with cohesive energy.\cite{Foppa-PRL-2022} This analysis illustrate how physical insights can be obtained from the key features identified by SGD. 

The rules associated to $SG_6^\mathrm{m}$ and $SG_{101}^\mathrm{m}$ are presented in the coordinates of the key parameters $a_0$ and $E_0$ in Fig.~\ref{fig:pos_mean}(D). We also present, in this figure, the rules associated to $SG_{61}^\mathrm{m}$, as an example of SG that belongs to the orange cluster of Fig.~\ref{fig:pos_mean}(B) whose rules only depend on $a_0$ and $E_0$ - see Table~\ref{tab:Rules}. In this plot, the bulk-modulus values are indicated by the grey scale color of the circles. The figure shows graphically that a more focused description is achieved as the utility-function values increases (and the SG size decreases) within the Pareto region. This figure also highlights that more focused rules might come at the expense of missing some high-bulk-modulus materials. For instance, some dark-grey circles corresponding to high-bulk-modulus materials are outside the limits of $SG_6^\mathrm{m}$. Thus, these high-bulk-modulus materials are not captured by $SG_6^\mathrm{m}$.

To understand which high-bulk-modulus materials might be "missed" by $SG_6^\mathrm{m}$ as an example of SG with high utility function, we verify whether the 5\% materials with the highest bulk moduli of the dataset are contained in $SG_6^\mathrm{m}$. These are 26 perovskites presenting bulk moduli higher than 1.43 eV/\AA$^3$ and composed by the $B$ elements chromium, manganese, iron, cobalt, and tungsten, and the $A$ elements scandium, praseodymium, neodymium, cerium, promethium, yttrium, samarium, and beryllium. 24 of these 26 materials satisfy the rules associated to $SG_6^\mathrm{m}$. The two high-bulk-modulus materials that do not satisfy the rules of $SG_6^\mathrm{m}$ are BeMnO$_3$ and BeWO$_3$, with bulk moduli of 1.43 and 1.45 eV/\AA$^3$, respectively. These two materials present the lowest cohesive energies (6.63 and 7.47 eV/atom, respectively) among the 26 materials. They are shown as lime crosses in Fig.~\ref{fig:pos_mean}(D). Thus, they do not satisfy the inequality $E_0>$7.48 eV/atom, which is part of the rules of $SG_6^\mathrm{m}$ (Table~\ref{tab:Rules}). The bulk modulus for these two materials could be governed by a different mechanism compared to the materials that are part of $SG_6^\mathrm{m}$. We will analyze this unexpected patter in more details in a dedicated subsection of the manuscript (see below).

\subsection*{Collection of SG Rules Obtained with the Cumulative-Jensen-Shannon-Divergence Utility Function}

We now turn our attention to the results obtained with an utility function based on the Jensen-Shannon divergence. The information-theoretic Jensen-Shannon divergence ($D_{JS}$) is a symmetrized version of the Kullback-Leibler divergence ($D_{KL}$), also known as relative entropy. The Jensen-Shannon divergence between the discrete distributions $R$ and $S$ is defined as

\begin{equation}
\label{eq:jsd}
\begin{split}
    D_{\mathrm{JS}}(R,S)&=\frac{1}{2} D_{KL}(R,M)
    +\frac{1}{2} D_{KL}(S,M)\\
    &=\frac{1}{2} \sum_{x \in \chi}R(x)log \left(\frac{R(x)}{M(x)}\right) \\
    &+\frac{1}{2} \sum_{x \in \chi}S(x)log\left(\frac{S(x)}{M(x)}\right),
\end{split}
\end{equation}
where $M(x)=\frac{R(x)+S(x)}{2}$, and $\chi$ indicates the sample space. 
The Jensen-Shannon divergence measures the dissimilarity between two distributions. It assumes small values for similar distributions and increases as the distributions are shifted with respect to each other. The value of the Jensen-Shannon divergence also increases if two distributions have different narrownesses. In our SGD analysis, the cumulative formulation of the Jensen-Shannon divergence,\cite{Nguyen-2015} denoted $D_{cJS}$, is used as utility function. The divergence is evaluated between the distribution of the target values in the SG and the distribution of the target values in the entire dataset. 
This utility function favors the selection of SGs presenting distributions of target values that are shifted and narrower compared to the distribution of the entire dataset. However, it does not explicitly requests that the values of the target in the SG are high or low.
Moreover, the cumulative Jensen-Shannon divergence does not make assumptions on the shape of the distributions. Thus, it can handle distributions that deviate significantly from a Gaussian more efficiently than utility functions based on the mean shift.
We stress that that other measures of similarity between distributions such as the Bhattacharyya distance can be used as utility functions in SGD. 

\begin{figure}[ht!]
\centering
\includegraphics[width=8cm]{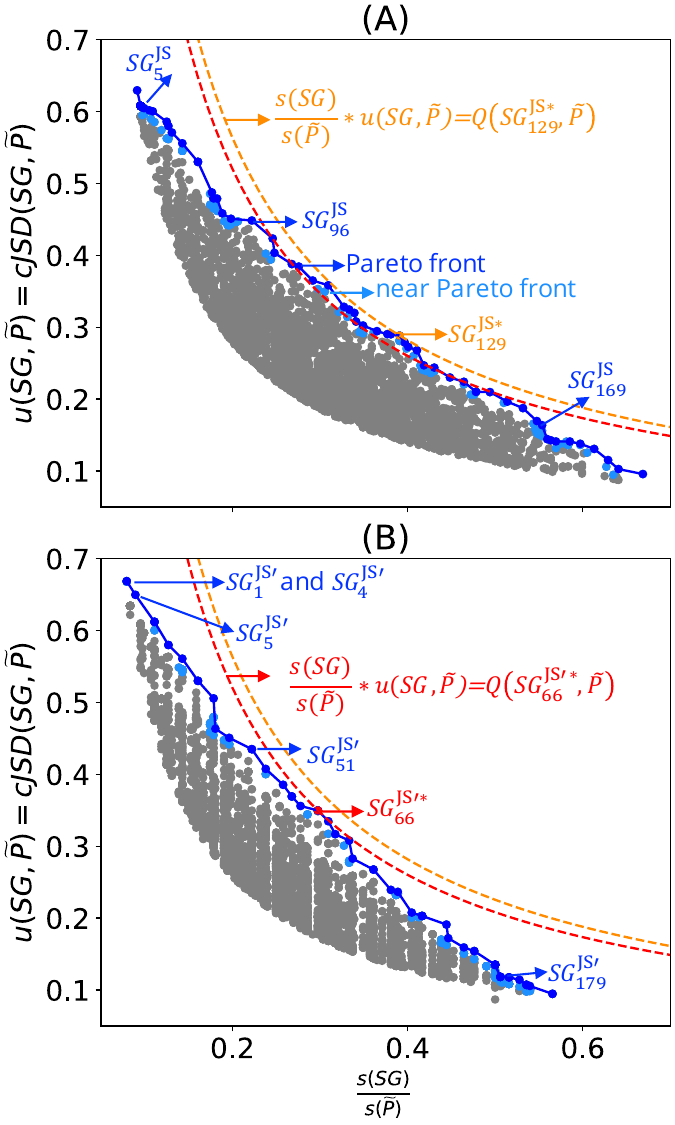}
\caption{Collections of SGs describing perovskites with high bulk modulus using the cumulative formulation of the Jensen-Shannon divergence as utility function. The 5,000 SGD solutions with high values of objective function are shown in grey and the Pareto region is displayed in blue. The SG associated to the maximum value of objective function is shown in orange or red. The two panels show results for two different sets of features. (A) Results for the full feature set containing the 24 features of Table I. (B) Results for the a reduced feature set containing 22 atomic and compositional features (see Table I). The rules associated to the SGs indicated in the figure are shown in Table II.}
\label{fig:cjsd}
\end{figure}

The results obtained with the cumulative Jensen-Shannon-divergence utility function and with the full set of 24 features are shown in Fig.~\ref{fig:cjsd} (A) and in Table~\ref{tab:Rules}. The Pareto front and region contain 101 and 189 SGs, respectively. The SGs identified at the Pareto region contain materials with high bulk modulus and they present narrow distributions of target values. 
We analyzed the selectors defining some of the SGs of this Pareto region. The rules associated with  $SG^{\mathrm{JS}}_{5}$, $SG^{\mathrm{JS}}_{96}$, $SG^{\mathrm{JS*}}_{129}$, and $SG^{\mathrm{JS}}_{169}$ constrain the values of the key features equilibrium lattice constant ($a_0$), cohesive energy ($E_0$), radius of valence-$s$ orbitals of +1 cations (cat) of $B$ element ($r_{s,B}^{\mathrm{cat}}$), the Kohn-Sham single-particle eigenvalue of the lowest-unoccupied orbital of the $B$ atom ($\epsilon_{\mathrm{L},B}$), the expected oxidation state of $B$ in the perovskite ($n_B$), the radius of the highest-occupied orbital of $A$ and $B$ neutral atoms ($r_{\mathrm{val},A}$ and $r_{\mathrm{val},B}$, respectively), and the electron affinity of the element $B$ ($EA_B$). Therefore, the rules highlight that the lattice constant and the cohesive energy are important parameters for describing high-bulk-modulus materials, along with atomic and compositional properties that mainly reflect the nature of the $B$ element of the $AB$O$_3$ perovskites.

Let us now compare the results obtained with the positive-mean-shift with the results obtained with the cumulative-Jensen-Shannon-divergence utility functions. The Pareto region of SGs identified based on the positive-mean-shift utility function is associated to relative SG sizes in the approximate range [0.20, 0.80] (Fig.~\ref{fig:pos_mean}(A)). The range of relative SG sizes in the Pareto region identified for the case of the cumulative-Jensen-Shannon-divergence utility function is, in turn, ca. [0.10, 0.70] (Fig.~\ref{fig:cjsd}(A)). Thus, optimal SGs with relative SG sizes below 0.20, i.e., SGs that contain less than 20\% of the dataset, could only be obtained with the the utility function based on the Jensen-Shannon divergence. These SG with small size are associated with distributions of bulk-modulus values that are narrower and more shifted towards high values than those associated to the SGs identified with the positive-mean-shift utility function. For instance, the materials selected by $SG^{\mathrm{JS}}_{5}$ and $SG^{\mathrm{JS}}_{96}$ present standard deviations of bulk-modulus values of 0.04 and 0.06 eV/\AA$^3$, respectively. The mean values of bulk modulus among the perovskites in these two SGs are 1.42 and 1.37 eV/\AA$^3$, respectively. None of the SGs identified with the positive mean shift presents higher mean values or lower standard deviation values (see Table~\ref{tab:Rules}). 
Thus, the rules corresponding to small SGs identified with the cumulative Jensen-Shannon divergence are more focused on high bulk modulus. This can be related to the fact that only this utility function explicitly favors narrow SGs. The SG rules identified using the cumulative-Jensen-Shannon-divergence utility function contain, in general, more statements and more features compared to the SG rules identified using the positive mean shift (Table~\ref{tab:Rules}). However, we note that the equilibrium lattice constant, the cohesive energy, and the radii of $B$ atoms are identified as key features by both approaches. 
The cumulative Jensen-Shannon-divergence utility function provides more focused rules for the present dataset and it is thus better than the positive-mean-shift for the purpose of identifying exceptional perovskites with very high bulk moduli. However, we stress that this utility function does not explicitly requests low or high values of target. This is a disadvantage compared to the positive-mean-shift utility function. Dispersion-corrected utility functions simultaneously take into account positive or negative shifts of the mean (or of the medians) of target values and the narrowness of the distributions of targets in the SG were proposed in order to simultaneously incorporate the requests for a shift in a specific direction and for small narrowness.\cite{Boley-2017}


\begin{table*}
\centering
\caption{Characteristics of Some of the SG Identified With the Pareto-Region Approach. Information on all other SGs of the Pareto Region is Provided in ESI. $^a$S, A, and C correspond to Solid, Atomic, and Compositional, respectively (See Table I). $^b$The star indicates the SG with the maximum value of objective (quality) function Q obtained with a given utility function and feature set. The superscripts "m" and "JS" correspond to the utility functions positive mean shift and cumulative Jensen-Shannon divergence, respectively. $^c$Standard deviation of the target within the SG.}
\label{tab:Rules}
\begin{tabular}{c c c c c c c l}
\toprule
Features & Index$^b$ & $Q(SG,\widetilde{P})$ & $u(SG,\widetilde{P})$ & $s(SG)/s(\widetilde{P})$ & $\bar{y}(SG)$ & $y_{\mathrm{std}}(SG)$$^c$ & Rules\\
\hline
S,A,C$^a$  & $SG_{1}^{\mathrm{m}}$ & 0.16 & 0.67 & 0.24 & 1.36 & 0.07 & $E_0 > 7.48~\mathrm{eV/atom} \wedge r_{s,B}^{\mathrm{cat}} \leq 1.44~\mathrm{\AA} $\\
\hline
S,A,C$^a$ & $SG_{6}^{\mathrm{m}}$ & 0.19 & 0.63 & 0.31 & 1.34 & 0.08 & $E_0 > 7.48~\mathrm{eV/atom} \wedge a_0 \leq 4.00~\mathrm{\AA} $\\
\hline
S,A,C$^a$ & $SG_{55}^{\mathrm{m*}}$ & 0.25 & 0.53 & 0.48 & 1.28 & 0.12 & $E_0 \geq 6.41~\mathrm{eV/atom} \wedge a_0 \leq 4.07~\mathrm{\AA} \wedge r_{s,B}^{\mathrm{cat}} \geq 0.94~\mathrm{\AA} $\\
\hline
S,A,C$^a$ & $SG_{61}^{\mathrm{m}}$ & 0.25 & 0.45 & 0.55 & 1.27 & 0.13 & $E_0 \geq 6.41~\mathrm{eV/atom} \wedge a_0 \leq 4.07~\mathrm{\AA}$\\
\hline
S,A,C$^a$  & $SG_{100}^{\mathrm{m}}$ & 0.20 & 0.26 & 0.77 & 1.19 & 0.18 & $E_0 \geq 5.42~\mathrm{eV/atom} \wedge a_0 \leq 4.16~\mathrm{\AA} $\\
\hline
S,A,C$^a$ & $SG_{5}^{\mathrm{JS}}$ & 0.06 & 0.61 & 0.10 & 1.42 & 0.04 & $-4.55 \leq \epsilon_{\mathrm{L},B} <-4.33 ~\mathrm{eV} \wedge  a_0 < 3.85~\mathrm{\AA} \wedge n_B < 3.5$  \\
\hline
S,A,C$^a$  & $SG_{96}^{\mathrm{JS}}$  & 0.10 & 0.45 & 0.22 & 1.37 & 0.06 & $ r_{\mathrm{val},A} \leq 1.51~\mathrm{\AA} \wedge EA_B \geq -1.84 ~\mathrm{eV} \wedge r_{s,B}^{\mathrm{cat}} \leq 1.50 ~\mathrm{\AA} $ \\
& & & & & & & $\wedge r_{\mathrm{val},B} \leq 1.14 ~\mathrm{\AA} \wedge E_0 > 7.11~ \mathrm{eV/atom} $ \\
\hline
S,A,C$^a$ & $SG_{129}^{\mathrm{JS*}}$& 0.11 & 0.29 & 0.39 & 1.32 & 0.09 & $ EA_B \leq 0.31~ \mathrm{eV} \wedge \epsilon_{\mathrm{L},B} \leq -3.50 ~\mathrm{eV} \wedge r_{s,B}^{\mathrm{cat}}  \geq 1.09 ~\mathrm{\AA} $  \\
& & & & & & & $\wedge E_0 \geq 6.77 ~\mathrm{eV/atom}$ \\
\hline
S,A,C$^a$ & $SG_{169}^{\mathrm{JS}}$ & 0.09 & 0.16 & 0.55 & 1.27 & 0.13 & $ E_0 \geq 6.41 ~\mathrm{eV/atom} \wedge a_0 \leq 4.07 ~\mathrm{\AA} $ \\
\hline
A,C$^a$  & $SG_{1}^{\mathrm{JS'}}$  & 0.05 & 0.67 & 0.08 & 1.43 & 0.03 & $EA_B \geq -1.03~\mathrm{eV} \wedge -4.55 \leq \epsilon_{\mathrm{L},B} < -4.33~\mathrm{eV} \wedge n_B<4 $ \\
\hline
A,C$^a$  & $SG_{4}^{\mathrm{JS'}}$  & 0.05 & 0.67 & 0.08 & 1.43 & 0.04 & $IP_B \leq 7.82~\mathrm{eV} \wedge \epsilon_{\mathrm{L},B} < -4.33~\mathrm{eV} \wedge  r_{\mathrm{val},B}<0.68~\mathrm{\AA}$ \\
& & & & & & & $\wedge n_B<3.5 $ \\
\hline
A,C$^a$  & $SG_{5}^{\mathrm{JS'}}$  & 0.06 & 0.65 & 0.09 & 1.42 & 0.04 & $ r_{\mathrm{val},A} < 1.28 ~\mathrm{\AA} \wedge EA_B \leq 0.31 ~\mathrm{eV} \wedge Z_B < 36 $  \\
& & & & & & & $\wedge r_{s,B} \geq 1.26 ~\mathrm{\AA}$ \\
\hline
A,C$^a$  & $SG_{51}^{\mathrm{JS'}}$ & 0.10 & 0.43 & 0.22 & 1.37 & 0.06 & $-5.11 \leq \epsilon_{\mathrm{L},B} \leq -3.50 ~\mathrm{eV} \wedge r_{\mathrm{val},B}^{\mathrm{cat}} \leq 0.94 ~\mathrm{\AA} \wedge n_A > 2 $  \\
\hline
A,C$^a$ & $SG_{66}^{\mathrm{JS'*}}$ & 0.10 & 0.35 & 0.30 & 1.34 & 0.08 & $r_{\mathrm{val},A}^\mathrm{cat} \leq 1.38~ \mathrm{\AA} \wedge \epsilon_{\mathrm{L},B} \leq -3.49 ~\mathrm{eV} \wedge r_{\mathrm{val},B}  \leq 1.14 ~\mathrm{\AA}$  \\
& & & & & & & $\wedge n_A \geq 1.5$\\
\hline
A,C$^a$ & $SG_{179}^{\mathrm{JS'}}$ & 0.06 & 0.12 & 0.52 & 1.26 & 0.15 & $ EN_A \geq 2.76 ~\mathrm{eV} \wedge EN_B \leq 4.85 ~\mathrm{eV} \wedge r_{s,B}  \geq 1.09 ~\mathrm{\AA}$ \\
\botrule
\end{tabular}
\end{table*}

\subsection*{SG Rules Obtained with a Reduced Set of Features}

So far, we have used the entire set of 24 features of Table 1 to obtain SGs of perovskites with high bulk modulus. SGD identified the equilibrium lattice constant ($a_0$) and the cohesive energy ($E_0$) among the key features required for describing high-bulk-modulus materials. However, in order to calculate $a_0$ and $E_0$ using DFT, the geometry of the materials needs to be optimized. This optimization corresponds to the majority of the work needed to calculate the bulk modulus itself. Thus, from the standpoint of exploring a large materials space, $a_0$ and $E_0$ are impractical (expensive) features, since one needs to evaluate these quantities for the materials under consideration in order to apply the SG rules. In order to obtain SG rules that describe high-bulk-modulus perovskites based on easily accessible features, we have also considered a reduced set of 22 features by excluding $a_0$ and $E_0$. Thus, we only considered the atomic and compositional features. In addition to this crucial cost aspect, this analysis also illustrates how the SG rules change when the feature set changes, and, in particular, when important features are not included in SGD. The identification of appropriate rules based solely on atomic and compositional features is a remarkable challenge for SGD, since the relationship between these basic features and the bulk modulus is significantly more indirect compared to the the relationship between bulk modulus and $a_0$ or $E_0$.\cite{Foppa-PRL-2022}     

We identified the Pareto region of SGs using the reduced set of 22 features and the cumulative Jensen-Shannon-divergence utility function. The SGs identifed with this approach are denoted $SG^{\mathrm{JS'}}$, where ' indicates the reduced feature set. The results are shown in Fig.~\ref{fig:cjsd}(B) and in Table~\ref{tab:Rules}. The identified Pareto front and near the Pareto front contain 66 and 136 SGs, respectively. The maximum value of the objective function obtained with the reduced feature set is slightly lower than that obtained with the full feature set (0.10 and 0.11, respectively, see orange and red dashed line in Fig. ~\ref{fig:cjsd}(A) and Fig. ~\ref{fig:cjsd}(B)). This indicates that the quality of the SG description is lower with the reduced feature set. The SG identified based on the reduced set of features that displays the maximum objective function, denoted $SG^{\mathrm{JS'}}_{66*}$ in Table~\ref{tab:Rules} and Fig.~\ref{fig:cjsd}(B), contains 150 materials. This SG contains less materials that the SG identified with all the features $SG^{\mathrm{JS}}_{129*}$ (197 materials). However, the overlap between the two SGs is significant. Indeed, 125 materials are present in both SGs. Thus, there is a high similarity between both descriptions. 

SG rules focusing on high-bulk-modulus materials at the low relative size portion of the Pareto region were obtained also with this reduced feature set. For instance, $SG^{\mathrm{JS'}}_{1}$, $SG^{\mathrm{JS'}}_{4}$, and $SG^{\mathrm{JS'}}_{5}$ display mean bulk-modulus values of 1.43, 1.43, and 1.42 eV/\AA$^3$ and standard deviations of 0.03, 0.04 and 0.04 eV/\AA$^3$, respectively. These figures are similar to those associated with the $SG^{\mathrm{JS}}_{5}$, which was identified based on the entire set of 24 features. The SGs $SG^{\mathrm{JS'}}_{5}$ and $SG^{\mathrm{JS}}_{5}$ contain, respectively, 45 and 48 materials. 40 materials are present in both SGs. This reflects a significant similarity between both descriptions. Thus, the SG with small relative size identified in the Pareto region with the reduced set of feature are comparable to those obtained with the full feature set.  

The rules associated with the SGs identified using the reduced set of 22 features depend on some key parameters that were also identified by the analysis of the entire set of 24 features and the cumulative-Jensen-Shannon-divergence utility function, e.g., radii of $A$ and $B$ elements, Kohn-Sham single-particle eigenvalue of the lowest-unoccupied orbital of the $B$ atom ($\epsilon_{\mathrm{L},B}$), electron affinity of $B$ atom ($EN_B$), and expected oxidation state of the $B$ element in the perovskite. However, some additional key features are identified in the analysis of the reduced set of features, such as the ionization potential of $B$ atom ($IP_B$), the atomic charge of $B$ element ($Z_B$), the expected oxidation state of the $A$ element in the perovskite ($n_A$), and the electronegativity of the $A$ element ($EN_A$). This shows how SGD attempts to reconstruct the information contained in the important features $a_0$ and $E_0$ by using the information on other offered features. Overall, the results obtained with the reduced feature set are comparable to those obtained with all 24 features. The rules derived based on the reduced feature set can thus be applied to identify high-bulk-modulus materials in larger materials spaces compared to the training set. 

\subsection*{Exploitation of the SG Rules for the Identification of Perovskites with High Bulk Modulus}
We applied the SG rules trained on 504 single $AB$O$_3$ perovskites with the reduced set of 22 features to identify materials with high bulk modulus ($B_0$) out of a candidate materials space of 12,096 compounds. This candidate material space was created by considering additional $A$ and $B$ elements that were not included in the training set ($A$: thorium and protactinium; $B$: hafnium, rhenium, osmium, iridium, gold, mercury, and thallium). Additionally, we combined two different $B$ elements to form double perovskites with the formula $A_2BB'$O$_6$. In the case of the double perovskites, the features related to the $B$ element are defined as the (composition) average of the features associated to the two different $B$ and $B'$ elements. In this analysis, we are explicitly considering a finite set of $A_2BB'$O$_6$ materials that contain a 1:1 $B$:$B'$ stoichiometric ratio. However, the material space of double perovskites is practically infinite, since any proportion of $B$ and $B'$ is possible, i.e., any $A_2B_{2-x}B'_x$O$_6$ formula with $x$ in the range [0.0,2.0] corresponds to a material in this space.    
In order to assess the usefulness of SG rules identified with the Pareto-region approach with respect to the SG rules associated to the maximum value of objective function, we applied two different sets of SG rules to select materials from the candidate materials space that likely present high bulk moduli. The first set of rules correspond to $SG^{\mathrm{JS'*}}_{55}$, which are associated to the SG with maximum value of the objective function. These rules are satisfied by 4,518 of the 12,096 candidates. The second set of rules correspond to $SG^{\mathrm{JS'}}_{1}$, $SG^{\mathrm{JS'}}_{4}$, and $SG^{\mathrm{JS'}}_{5}$, which are associated to the SGs at the Pareto region with high utility-function values, i.e., high exceptionality. These rules are satisfied by 238 materials, out of the 12,096 candidates, i.e., 1.97\% of the candidate materials space. Then,  we randomly selected 50 of the 4,518 selected materials and 50 of the 238 selected materials and evaluated their $B_0$ by DFT-PBEsol calculations. For comparison, we have also calculated the $B_0$ of 50 perovskites that were randomly selected from the 12,096 materials. 

\begin{figure}[ht!]
\centering
\includegraphics[width=8cm]{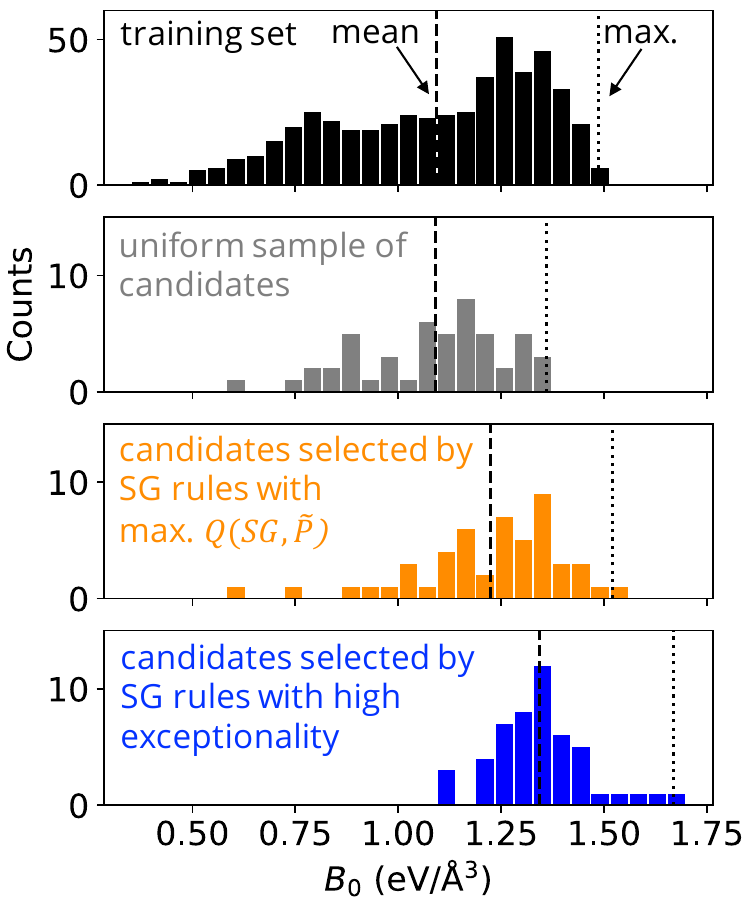}
\caption{The SG rules describing perovskites with high bulk modulus ($B_0$) trained on 504 single $AB$O$_3$ perovskites are applied to identify promising single $AB$O$_3$ and double $A_2BB'$O$_6$ perovskite from a candidate space containing 12,096 materials. The histograms show the distribution of $B_0$ among the materials of the training dataset (in black), among 50 materials randomly selected from the candidate space (in grey), among 50 materials of the candidate space selected according to the SG rules $SG^{\mathrm{JS'*}}_{66}$, and among 50 materials of the candidate space suggested by the SG rules $SG^{\mathrm{JS'}}_{1}$, $SG^{\mathrm{JS'}}_{4}$, and $SG^{\mathrm{JS'}}_{5}$ (bottom, in blue). The dashed and dotted lines indicate the mean and maximum (max.) $B_0$ values of each distribution, respectively.}
\label{fig:screening}
\end{figure}

The distribution of $B_0$ for the materials that were randomly selected from the candidate materials space (Fig.~\ref{fig:screening}, in grey) has practically the same mean value compared to that of the distribution of $B_0$ in the training set (Fig.~\ref{fig:screening}, in black), i.e., 1.09 eV/\AA$^3$. This indicates that the training data might be representative of this specific candidate materials space. The highest $B_0$ among the materials randomly selected from the candidate materials space is 1.36 eV/\AA$^3$. This value is significantly lower than the highest $B_0$ in the training dataset, 1.49 eV/\AA$^3$ for ScMnO$_3$.

The distribution of $B_0$ for the materials that were selected from the candidate materials space by the SG rules $SG^{\mathrm{JS'*}}_{55}$ (Fig.~\ref{fig:screening}, in orange) is concentrated on high $B_0$, with a mean $B_0$ value of 1.22 eV /\AA$^3$. One material suggested by theses SG rules has $B_0$ higher than the highest value in the training dataset, PaCoOsO$_6$, with bulk modulus of 1.52 eV /\AA$^3$, respectively. This value is slightly higher than the highest $B_0$ in the training dataset (1.49 eV/\AA$^3$). However, we note that the rules suggested several materials that turned out to have relatively low values of $B_0$.

The distribution of $B_0$ for the materials that were selected from the candidate materials space by the SG rules $SG^{\mathrm{JS'}}_{1}$, $SG^{\mathrm{JS'}}_{4}$, and $SG^{\mathrm{JS'}}_{5}$ (Fig.~\ref{fig:screening}, in blue) is concentrated on higher values compared to the other distributions, with a mean $B_0$ value of 1.35 eV /\AA$^3$. Additionally, four of the materials suggested by the SG rules have $B_0$ higher than the highest value in the training dataset. These are PaMnO$_3$, Pa$_2$CrFeO$_6$, Pa$_2$VCrO$_6$, and PaVO$_3$, with bulk modulus of 1.67, 1.63, 1.59, and 1.55 eV /\AA$^3$, respectively. Thus, the SG rules lead to the identification of materials with $B_0$ up to 13\% higher than any observation of the training dataset. This result indicates that the SGs identified by the Pareto-region analysis can point at more exceptional materials compared to the standard SGD approach. In particular, in the present use-case, materials that present higher performance than the known compounds of the training set were more efficiently selected by the focused SG rules provided by the multi-objective optimization of SGs.  

The dataset utilized to train SGD and the verification of SGD suggestions are based on DFT-PBEsol calculations. Even though DFT-PBEsol presents an overall good accuracy for describing properties of solid materials such as the bulk modulus,\cite{Zhang-2018} the errors of DFT-PBEsol should be taken into account when comparing the reported bulk moduli with experimental results. 

\subsection*{Investigation of High-Bulk-Modulus Perovskites that Are Not Captured by the SG Rules}

When analyzing the SG rules, we highlighted that  BeMnO$_3$ and BeWO$_3$ present bulk moduli above the 95\%-ile of the bulk-modulus distribution in the training set (1.43 and 1.45 eV/\AA$^3$, respectively) but they are not contained in the high-utility-function SG $SG_6^\mathrm{m}$ identified in Fig.~\ref{fig:pos_mean}. 
In the perovskite structure, the $A$ cations are larger than the $B$ cations and the relationships between the radii of the $A$ and $B$ cations and the anions determine the thermodynamic stability of the perovskite structure, as given by the tolerance factors.\cite{Goldschmidt-1926, Bartel-2019} Beryllium and magnesium are the only two $A$ elements in our dataset for which the radius of $A$ (e.g., $r_{s,A}$) might be smaller than the radius of $B$ (e.g., $r_{s,B}$). This is the case for the materials BeMnO$_3$ and BeWO$_3$. Indeed, $r_{s,\mathrm{Be}}<r_{s,\mathrm{Mn}}$ and $r_{s,\mathrm{Be}}<r_{s,\mathrm{W}}$. Thus, beryllium would most likely occupy the $B$ site of these cubic perovskite structures. Indeed, several reports discuss the properties of perovskites composed by beryllium at the $B$ sites, i.e., coordinated with 6 oxygen or halide anions.\cite{Li-2008,Mahmood-2019,Kumari-2024} This observation motivated us to evaluate the bulk modulus of the perovskites MnBeO$_3$ and WBeO$_3$, where beryllium sits at the $B$ site and manganese or tungsten sit at the $A$ sites. The calculated bulk modulus are equal to 1.44 and 1.82 eV/\AA$^3$. The bulk modulus of WBeO$_3$ is 22\% higher than the highest value in the training set and is the highest bulk modulus identified in this paper. We have also evaluated the bulk modulus of perovskites with $B$=Be and other transition metals of the third row of the periodic table as $A$ elements, namely hafnium, tantalum, rhenium, osmium and iridium. The bulk modulus of the materials HfBeO$_3$, TaBeO$_3$, ReBeO$_3$, OsBeO$_3$, and IrBeO$_3$ are equal to 1.70, 1.81, 1.75, 1.63, and 1.57 eV/\AA$^3$, respectively. These values are also relatively high compared to the values of the training set and they show that several beryllium based materials might be exceptional.     
This analysis illustrates how the exploratory nature of SGD analysis can identify unexpected patterns and anomalies, which might lead in turn to the identification of exceptional materials. 

We note that the rules identified with SGD will be valid as long as the physical processes governing the materials in the training dataset also govern the behavior of the materials in the materials space to be explored. In order to cover portions of the materials space where different underlying processes from those present in the training set are important, the incorporation of new data points and retraining of SG rules will be required. Finally, useful materials might present unusual combinations of different materials properties. These could also be consider exceptional materials. Identifying such materials calls for multi-objective optimization of materials properties.\cite{DelRosario-2020,Jablonka-2021,Shi-2023,Low-2024} The SGD can be adapted for this scenario and this aspect will be addressed in an upcoming contribution.

\section{Conclusion}
We introduced an approach for the identification of coherent collections of SGs of the "Pareto region" with respect to the SG size and exceptionality objectives of the SGD analysis. The concept was demonstrated by the learning of rules that describe perovskites with high bulk modulus. Our results show that rules focused on exceptional materials do not necessarily correspond to the one SG that maximizes the objective function, but these rules can be identified with the Pareto-region concept. This analysis does not require additional computational effort, since the SGD solutions with high objective-function values are obtained on the fly during the optimization of the objective function. We used the SG rules obtained by the multi-objective approach to identify exceptional perovskites with bulk modulus up to 13\% higher than the highest value contained in the training set of 504 materials, out of a materials space of more than 12,000 materials.

\section{Methods}
\textbf{SGD Approach} We used the SGD approach implemented in the realkd version 0.7.2. A Monte-Carlo-based SG search algorithm\cite{Boley-2011,Boley-2012} was used with 50,000 seeds for the initialization. The propositions were created based on 10 different thresholds (or cutoffs) per feature. These thresholds were determined based on $k$-means clustering. 
For a desired number of $k$ threshold values, $k+1$ clusters are obtained. Then, thresholds are taken as the arithmetic means between the maximum and minimum values of a given feature in neighboring clusters.

\section{Data Availability}
The dataset, including all input and output files of the calculations, are available in reference \cite{Foppa-PRL-2022}. 

\begin{acknowledgments}
This work was funded by the NOMAD Center of Excellence (European Union’s Horizon 2020 research and innovation program, Grant Agreement No. 951786) and the ERC Advanced Grant TEC1p (European Research Council, Grant Agreement No 740233).
\end{acknowledgments}
*foppa@fhi-berlin.mpg.de
\bibliography{main}
\bibliographystyle{unsrt}

\end{document}